\documentclass[a4paper,11pt]{article}
\pdfoutput=1 % if your are submitting a pdflatex (i.e. if you have
\usepackage{jinstpub} % for details on the use of the package, please
\usepackage{subcaption} 
 
\title{\boldmath Using Machine Learning to Speed Up and Improve Calorimeter R\&D}

\author{F. Ratnikov}
\affiliation{NRU Higher School of Economics,\\Moscow, Russia}

\emailAdd{fedor.ratnikov@cern.ch}

\abstract{
Design of new experiments, as well as upgrade of ongoing ones, is a
continuous process in the experimental high energy physics. 
Frontier R\&Ds are used to squeeze the maximum physics performance using cutting edge detector technologies.
The evaluation of physics performance for particular configuration
includes sketching this configuration in Geant,  simulating typical
signals and backgrounds, applying reasonable reconstruction
procedures, combining results in physics performance metrics.
Since the best solution is a trade-off between different kinds of
limitations, a quick turn over is necessary 
to evaluate physics performance for different techniques in different configurations.
Two typical problems which slow down evaluation of physics performance
for particular approaches to calorimeter detector technologies and
configurations are: 
\begin{itemize}
\item Emulating particular detector properties including raw detector
  response together with a signal processing chain to adequately
  simulate a calorimeter response for different signal and background
  conditions. This includes combining detector properties obtained from the general Geant simulation with properties obtained from different kinds of bench and beam tests of detector and electronics prototypes. 
\item Building an adequate reconstruction algorithm for physics
  reconstruction of the detector response which is reasonably tuned 
to extract the most of the performance provided by the given detector
configuration.
\end{itemize}

Being approached from the first principles, both problems require
significant development efforts. Fortunately, both problems may be
addressed by using modern machine learning approaches, that allow a
combination of available details of the detector techniques into 
corresponding higher level physics performance in a semi-automated way.

In this paper, we discuss the use of advanced machine learning techniques to speed up and improve the precision of the detector development and optimisation cycle, with an emphasis on the experience and practical results obtained by applying this approach to epitomising the electromagnetic calorimeter design as a part of the upgrade project for the LHCb detector at LHC.
}


\collaboration[c]{on behalf of the LHCb Calorimeter Upgrade group}

\proceeding{Calorimetry for High Energy Frontier\\
  November 25-29, 2019\\
  Fukuoka, Japan}

\begin{document}
\maketitle
\flushbottom

\section{Introduction}

Design of new experiments, as well as upgrade of ongoing experiments, is a
continuous process in experimental high energy physics.

This is a many-fold process: global optimisation requires different steps to be coordinated. For example, when varying the material of the calorimeter absorber, the reconstruction algorithm should be re-tuned to accommodate a new Moli\`ere radius for the detector. If done manually, this is a time-consuming procedure, which significantly slows down the optimisation loop turnover.

 The ultimate goal of the detector
construction is to establish the necessary performance of physics measurements.
However, the propagation of particular future calorimeter technologies to the  metrics quantifying the ultimate physics performance of the detector is not immediate, and requires several steps in between.

\begin{figure}[htbp]
\centering 
\includegraphics[width=0.8\textwidth]{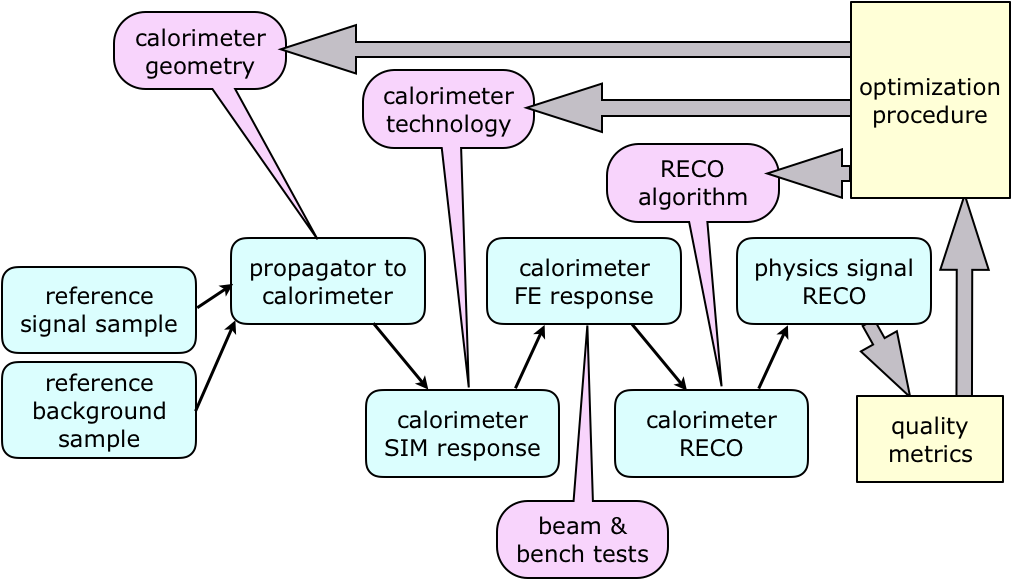}
\caption{\label{fig:pipeline} General pipeline for the calorimeter
  optimisation includes several steps. Blue blocks indicate data processing pipeline steps; pink bubbles represent configurations and conditions for pipeline steps; yellow blocks close down the optimisation loop.}
\end{figure}

The typical workflow for optimisation of the calorimeter components is
sketched in Fig.~\ref{fig:pipeline}.
\begin{itemize}
\item Selected event  samples, both signal and background, are used to
  initiate an optimisation cycle for comparing performance for signal
  recovery and background suppression.
\item Calorimeter is usually installed downstream of the detector, so propagation of
  events from the origin point to the calorimeter is necessary. This step is dependent on 
  the properties of the rest of the detector in front of the
  calorimeter. Also, if the calorimeter detector has an inhomogenous
  configuration, details of the global geometry are to be accounted for in this
  step. The latter allows  optimisation of the global geometry based on
  the physics quality metrics.
\item The technology for the individual calorimeter modules is a central point for the detector R\&D. To evaluate technology, we need to simulate its response to the event. This is done using response simulation models, e.g. Geant4~\cite{geant4}. The calorimeter technology details drive such simulation.
\item Behaviours of the front end electronics are another important contribution into physics quality of the detector. Although such properties are hard to simulate, good data samples may be obtained from beam or bench tests.
\item The reconstruction algorithm is absolutely necessary to evaluate the quality of converting the detector response into the physics objects.
\item The physics quality metrics may be calculated using reconstructed objects. This metric can be used as a target function for the optimisation procedure. 
\item All aspects of the calorimeter may be optimised: the details of the calorimeter technology, the geometrical layout, and possible reconstruction algorithms.
\end{itemize}

Optimisation cycle built on top of event processing pipeline allows to derive physics motivated optimal parameters for the detector.   

\section{Machine Learning Based Approach}

To evaluate the physics performance of a particular configuration of the possible future calorimeter detector, one needs to run the optimisation cycle described above. A good fine-tuning of individual blocks is important to properly propagate properties of the configuration under study to the ultimate physics performance. For regular stable detector operation these blocks are carefully tuned for the actual detector configuration. In contrast, for the detector R\&D process, many different possible configurations are studied simultaneously. 
Nevertheless, the reasonable representations of simulation and reconstruction steps, which are tuned for every studied configuration, are necessary for inferring consistent conclusions about physics performance of these configurations. Being made manually, this is a time consuming work. Fortunately, these studies use well labelled datasets either from MC simulation or from test beam measurements. Thus surrogate models may be built and trained on labelled data using regular Machine Learning (ML) approaches. This allows to speed up building models for different pipeline steps. Importantly, such training may be automated and requires minor expert supervision.
In the following sections we demonstrate the possibility of applying ML based solutions to different steps of the pipeline.

\section{Generating Detector Response}

\textsc{Geant4} simulation of the calorimeter response is computationally intensive. This is primarily because shower development and collecting responses are done on the micro-level of the individual shower particles. At the same time, calorimeters usually have granularity much coarser than that of the simulation micro-level. As a result, detailed shower information is aggregated into relatively few responses of the calorimeter. This means that the transfer function converting impact particle parameters into the calorimeter responses is relatively simple and maybe substituted by the surrogate generative model trained by standard means of ML. 

\begin{figure}[htbp]
\centering 
\includegraphics[width=0.8\textwidth]{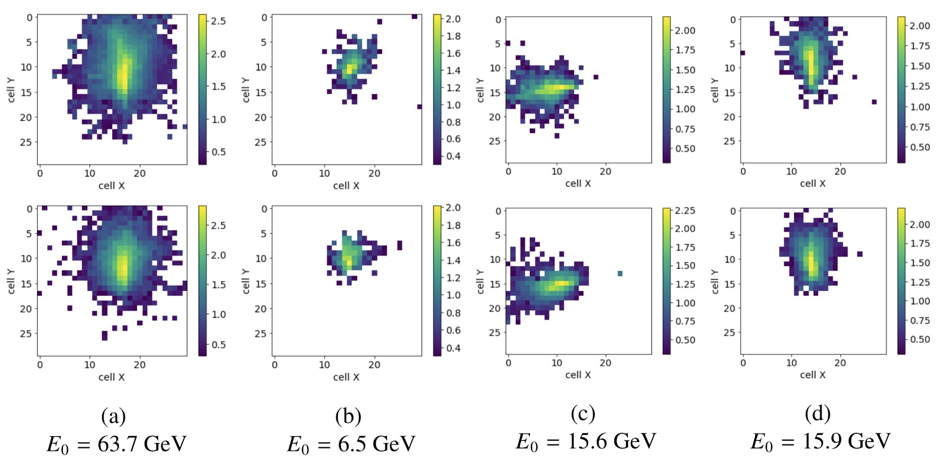}
\caption{\label{fig:calogan} Using generative models may
  significantly speed up simulation of detector response. Top row: Geant4 simulated showers; bottom row: generated calorimeter responses for impact particle identical to those producing showers in the top row~\cite{lhcbcalogan}.}
\end{figure}

Fig.~\ref{fig:calogan} illustrates application of the surrogate generative model to the LHCb electromagnetic calorimeter~\cite{lhcbcalogan}. While the calorimeter has modules with three different granularities of 12, 6, and 4~cm in size, the generative model based on the Wasserstein flavor~\cite{wsgan} of Generative Adversarial Network approach~\cite{gan} converts kinematics and position of the impact particle into virtual calorimeter response on the 30x30 matrix of cells of 2~cm in size. This allows to aggregate the obtained response into every actual calorimeter granularity around the impact point. Details in Ref.~\cite{lhcbcalogan} demonstrate that such a model easily learns main properties of the signal thus allowing 3 orders of magnitude faster simulation of the calorimeter response for the R\&D optimisation cycle. 

\section{ML-based Reconstruction}

The big slow down factor for running optimisation cycle is a necessity to fine tune reconstruction algorithm for every new calorimeter technology and geometry configuration. ML may help to tune the reconstruction in an automatic way. Indeed, as soon as a sample of calorimeter responses is available, the corresponding regressor may be trained to extract physics information from the raw response.

\begin{figure}[htbp]
\centering 
\includegraphics[width=0.8\textwidth]{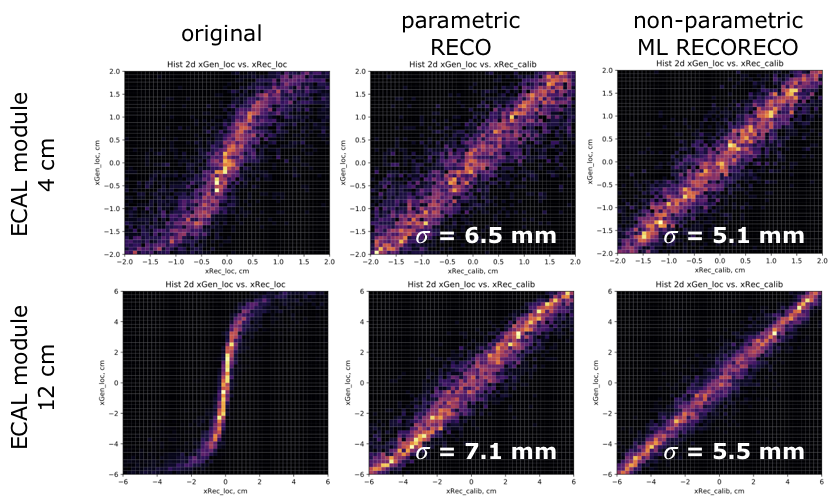}
\caption{\label{fig:scurve} ML-based reconstruction of the calorimeter
  cluster position provides spatial resolution similar to the customised
  reconstruction procedure, but without {\em a priori} knowledge about
  the particular spatial properties of the calorimeter under study. Left - correlation between cluster centre and the true track position; middle - correlation corrected using parametrised correction; 
right - correlation using ML trained regressor.  }
\end{figure}

Fig.~\ref{fig:scurve} demonstrates the quality of the spatial reconstruction of the calorimeter cluster for the case of the LHCb 4~cm and 12~cm modules. Agnostic to particular calorimeter details, automatically trained ML model (based on xgboost~\cite{xgboost} in this case) produces a slightly better performance than the manually selected parametric model. Importantly, the automatically trained generic regressor provides performance comparable with the manually tuned one. This justifies the use of this regressor in the optimisation cycle in place of a well tuned reconstruction algorithm for extracting physics observables from the calorimeter response.

\section{Pileup Mitigation with Timing}
A significant contribution from pileup events is expected after LHC and detectors upgrade during long shutdown in 2031~\cite{lhcls3}. The effective way of reducing this contribution is using signal time, thus separating contributions from distinct primary interactions. To evaluate this approach, consistent estimation of timing resolution for different technologies and configurations of the upgraded calorimeter are required. Using ML for this task allows making this estimation automatic, consistent, and agnostic to details of the system.

For this analysis we use responses from 30 GeV electrons collected at DESY test beam facility~\cite{testbeamdata}. Signals are read out at a sampling rate of 5~GHz, which allows re-sampling to lower the effective sampling rates. Details of data and its pre-processing are described in Ref.~\cite{chepposter}.

\begin{figure}[htbp]
\centering 
\begin{subfigure}{0.58\textwidth}
 \centering
\includegraphics[width=1.\textwidth]{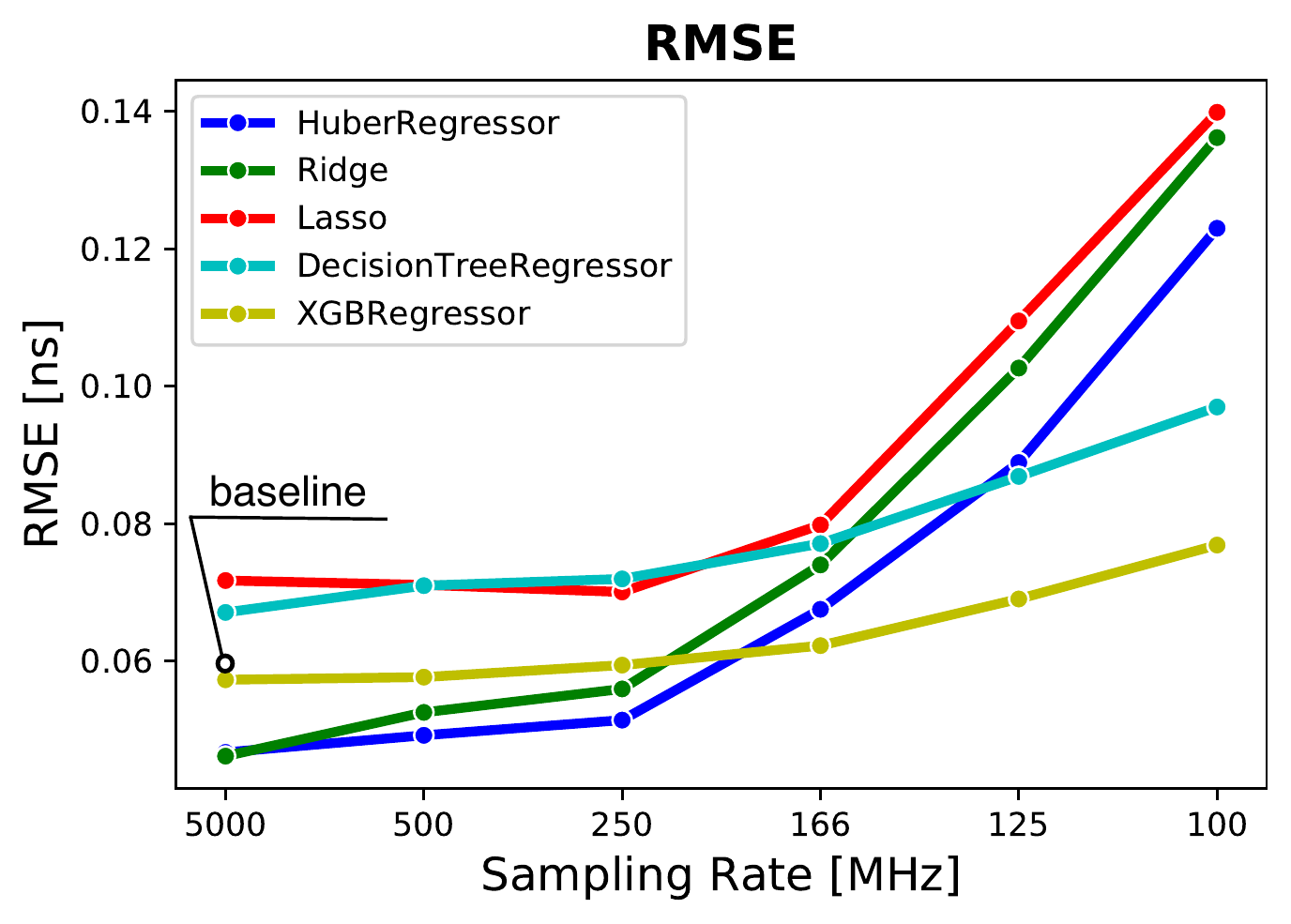}
 \caption{\label{fig:timeresolution} Time resolution obtained for different ML based regressors for different sampling rates. 
  Baseline estimates time of the signal as a half height of the signal front.}
\end{subfigure}
\hfill
\begin{subfigure}{0.38\textwidth}
\centering 
\includegraphics[width=1.0\textwidth]{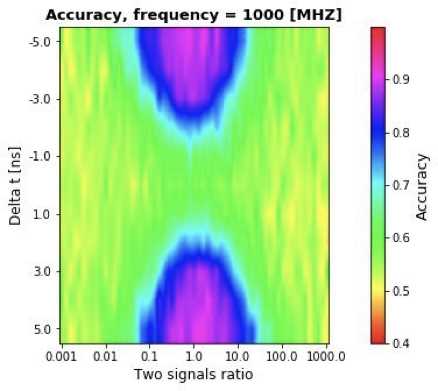}
\caption{\label{fig:onetwodisc} Accuracy for the classifier which distinguishes one-pulse signal and composition of one primary signal and one background signal.
  There is a well defined phase space where signal and composite signals may be separated.}
\end{subfigure}
\caption{}
\end{figure}

 ML regressor may be trained on test beam data to convert a set of signal sampling series into the reference time of the signal.
 To test stability of the ML approach, several different ML techniques are used to train the regressor. Fig.~\ref{fig:timeresolution} demonstrates similar best results for different sampling rates obtained from several different approaches. This consistency confirms that different flexible enough surrogate ML models reproduce time measurement behaviours with reasonable accuracy, and it is good enough for approximate R\&D studies. The result illustrates that for these kinds of signals using sampling rates above 500~MHz insignificantly improves timing resolution.
 The xgboost ML implementation is used in following examples as the most common, flexible and stable approach.   

At high pileup conditions, there is a possibility that the same channel will contain smaller background contribution from another signal in addition to the primary signal. That background contribution will have a different time, and thus may disturb the timing measurement for the primary signal.
To evaluate the efficiency of the pileup mitigation, it is important to estimate this effect. This requires a customised reconstruction algorithm, that accounts for details of signals behaviours. 
The ML approach to the same problem is generic: it does not require {\em a priori} knowledge details of signal properties. All the necessary information is extracted from the train data sample.  
Following this approach, the dataset was prepared by constructing composite signals as an artificial mixture of the primary and the background signals with known amplitude ratio and time offset.
To evaluate the ability to detect the presence of another, background contribution, the ML classifier is trained on this dataset. The ML classification quality for identifying the presence of  background contribution  for different relative signal strengths and time offsets is presented in Fig.~\ref{fig:onetwodisc}.

\begin{figure}[htbp]
\centering 
\begin{subfigure}{0.44\textwidth}
 \centering
\includegraphics[width=1.\textwidth]{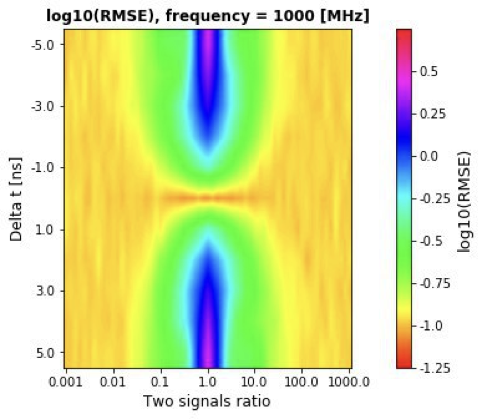}
\caption{\label{fig:timeresolution2} Timing resolution of the primary signal in presence of the background signal for the readout sampling rate of 1000 MHz. 
There is a well defined phase space where background contribution significantly disturbs timing measurements.}
\end{subfigure}
\hfill
\begin{subfigure}{0.52\textwidth}
 \centering
\includegraphics[width=1.\textwidth]{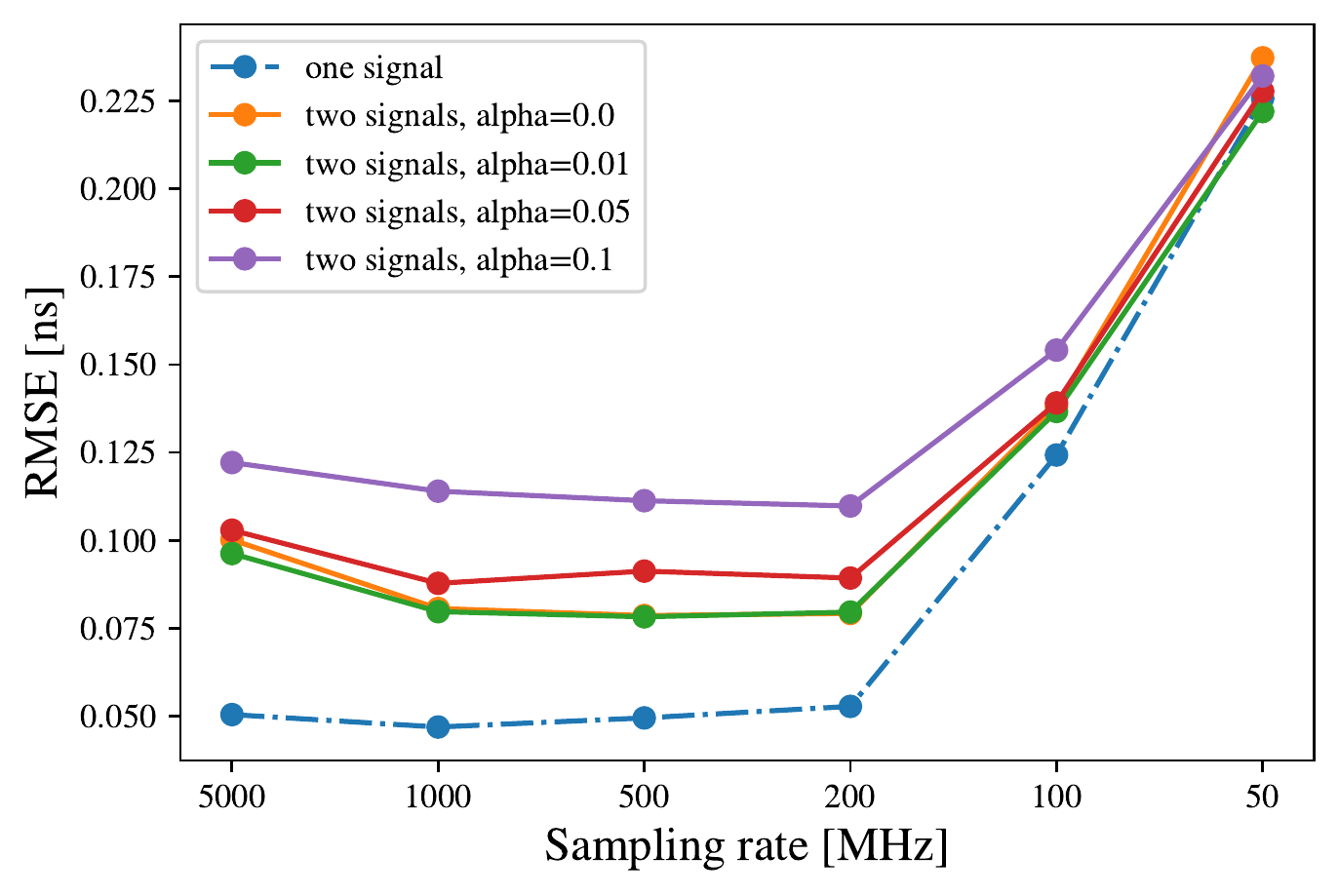}
\caption{\label{fig:timeresolution1}Timing resolution of the primary signal in presence of the background signal as a function of sampling rate for different relative amplitudes of the background signal.
Time offset is assumed normally distributed with $\sigma$=1~ns.}
\end{subfigure}
\caption{}
\end{figure}

The next problem is to quantify the effect of the background contribution on the time resolution of the primary signal. The regressor, which converts composed signal time series to the reference time of the primary signal, was trained using a set of composite signals. Fig.~\ref{fig:timeresolution2} illustrates time resolution degradation for different relative amplitudes and time offsets of the two 
contributions. This distribution is necessary to plug time resolution effects into the signal reconstruction step of the optimisation loop. Fig.~\ref{fig:timeresolution1} illustrates time resolution dependency on the sampling rate. The difference between performances for one signal and for two signals including background contribution $\alpha$=0 is driven by insensitive region in Fig.~\ref{fig:onetwodisc} where the regressor can not reliably identify the presence of the second contribution. The {\em a priori} knowledge that it is a single contribution thus improves the timing resolution for the signal.

\section {Global Optimisation}

The pipeline in Fig.~\ref{fig:pipeline} with plugged in ML driven steps allows automation of the optimisation cycle. Although the automatic training of ML components is much faster than manual tuning,
it still takes significant time. Therefore, the global optimisation procedure requires minimising the total number of optimisation iterations . ML suggests effective approaches to this kind of optimisation problems, like Bayesian optimisation and others~\cite{mloptimisation}. ML based multi-parameter optimisation with a massive Geant4 simulation step included into the optimisation cycle is illustrated in Ref.~\cite{shipopt}.

\section{Conclusions}

The calorimeter R\&D process requires time consuming computation steps to evaluate physics performance for different detector techniques and configurations. Surrogate ML models may be used for most steps that are necessary for evaluating quality of different solutions. Such models are automatically trained on available datasets and provide possibility to consistently estimate the resulting physics performance. Using automatic training speeds up the turnover for the performance studies and ensures consistency and uniformity of obtained results.   

\acknowledgments

The research leading to these results has received funding from the
Russian Science Foundation under agreement No 19-71-30020.


\end{document}